\documentclass[12pt,a4paper]{article}
\usepackage{epsfig}
\usepackage{color}
\usepackage{graphicx}
\usepackage{amsmath,amssymb}
\usepackage{authblk}

\def\plb#1#2#3{Phys. Lett. B {\bf #1}, #2 (#3)}

\def\prd#1#2#3{Phys. Rev. D {\bf #1}, #2 (#3)}

\begin{document}
\date{}
%\begin{frontmatter}
\title{{\large\bf Hadron Production in Ultra-relativistic
Nuclear Collisions and the QCD Phase Diagram: an Update}}

\author[1,2,3,4]{P.\ Braun-Munzinger}
\author[5] {J.\ Stachel}

\affil[1]{GSI Helmholtzzentrum f\"ur Schwerionenforschung,
  D-64291 Darmstadt, Germany}
\affil[2]{ExtreMe Matter Institute EMMI, GSI, D-64291 Darmstadt,
  Germany}
\affil[3]{Technische Universit\"at Darmstadt, D-64289 Darmstadt,
  Germany}
\affil[4]{Frankfurt Institute for Advanced Studies, J.W.\ Goethe
  Universit\"at, D-60438 Frankfurt, Germany}
\affil[5]{Physikalisches Institut der Universit\"at
  Heidelberg, D-69120 Heidelberg, Germany}

\maketitle

\begin{abstract} 
We summarize our current understanding of the connection between the
QCD phase line and the chemical freeze-out curve as deduced from
thermal analyses of yields of particles produced in central collisions
between relativistic nuclei.

\end{abstract}

%\begin{keyword} Dense quark matter, chemical freeze-out, QCD phase diagram
%\PACS{12.39.Fe, 11.15.Pg, 21.65.Qr} \end{keyword}

%\end{frontmatter}

%%%%%%%%%%%%%%%%%%%%%%%%%%%%%%%%%%%%%%%%%%%%%%%%%%%%%%%%%%%%%%%%
%%%%%%%%%%%%%%%%%%%%%%%%%%%%%%%%%%%%%%%%%%%%%%%%%%%%%%%%%%%%%%%%
%\section{Introduction}
\vspace{5.0mm}
Quantum chromodynamics (QCD), the theory of strong interactions, contains, as
a key prediction, a new phase of matter, the quark-gluon plasma (QGP). This
partonic QGP was the main ingredient (in addition to leptons and neutrinos) of
the matter in the early universe until about 10 $\mu$s after the big
bang. Around that time the QGP was transformed into hadronic matter through
the QCD phase transition. In general, strongly interacting matter in
equilibrium is characterized by two quantities, the temperature $T$ and the
baryon chemical potential $\mu_B$ or equivalently, the (net) baryon density
$n_B$. 

How does this phase transition come about? Due to asymptotic freedom, the
strong coupling constant $\alpha_s$ runs, i.e. diminishes with increasing
energy scale, implying that interactions among strongly interacting particles
will get weaker as $T$ or $\mu_B$ (or both) are increased. One may ask at what
energy scale will there be new physics such as deconfinement for strongly
interacting matter in equilibrium?  The only two scales relevant appear to be
$\Lambda_{QCD} \approx 200$ MeV along the temperature axis and the nucleon
mass $m_N$ along the $\mu_B$ axis. From phenomenological considerations and
based on ever more accurate solutions of QCD on a discrete space-time lattice
('lattice-QCD') convincing arguments were put forward to demonstrate that
strongly interacting matter undergoes indeed a (phase) transition from a dense
hadronic medium where all constituents (hadrons) are confined to a deconfined
plasma of interacting quarks and gluons. This implies the existence of a line
in the ($T - \mu_B$) plane, the QCD phase boundary, anchored by parametrically
critical parameters $T_c(\mu_B =0) \approx \Lambda_{QCD}$ and $(\mu_B)_c(T=0)
\approx m_N$. For recent detailed reviews on the QCD phase dia\-gram, where
many of the above short arguments are exposed in detail and the state of
theory and relevant experiments is summarized, see
\cite{pbm_wambach,fukushima_hatsuda}.

This article is an attempt to update a paper on the subject which we
wrote 15 years ago \cite{pbm_js_gerry70} in honor of Gerry Brown's
70th birthday. While we have (with one exception \cite{welke}) never
published together with Gerry, his constant probing and provocative
questions on the QGP and related areas have, during nearly 2 decades
of close scientific interactions, deeply influenced our thinking on
the subject. The experimental basis for making a connection between
data from ultra-relativistic nuclear collisions and the QCD phase
boundary is dramatically improved after 15 years of intense research
on the subject. Given all this we consider it an honor and an
opportunity to be able to summarize our most recent thinking on the
subject as part of a Festschrift on the occasion of Gerry's 85th
birthday.

Strongly interacting matter at high temperature and baryon density is
produced in collisions between atomic nuclei at ultra-relativistic
energies. Depending on the center-of-mass energy reached in the
collisions the temperature and baryon density can be 'tuned'
systematically, as discussed in detail below. A sequence of such
experiments conducted over the past 25 years over a wide range of
energies has provided convincing evidence for a new state of matter,
see, e.g., \cite{Specht:2001qe,Heinz:2000bk,
  Stachel:1998rc,Arsene:2004fa,Back:2004je,Adams:2005dq,Adcox:2004mh,
  BraunMunzinger:2007zz}. In November 2010, first results from Pb--Pb
collisions at the Large Hadron Collider at CERN have provided a first
glimpse into fireballs with ultra-high temperatures (approaching 1
GeV) and opened a new era in quark matter research
\cite{alice_pb1,alice_pb2,alice_pb3,alice_pb4,alice_pb5,atlas}.

The range of temperatures and densities reached in these experiments
is clearly of the order of or (for RHIC and LHC experiments)
significantly exceeding the typical values discussed above. For a
review and summary of the evidence, see
\cite{gyulassy_mclerran}. Much harder has been to assess the
degree of equilibration reached in the collisions and to provide
direct evidence on the QCD phase boundary. Below we will argue that
the most direct information on equilibration and the QCD phase
boundary available to-date is obtained from analyses of the
multiplicities of hadrons produced in central collisions between
ultra-relativistic nuclei.

\begin{figure}[htb]
\begin{center}
\resizebox*{!}{8.0cm}{\includegraphics{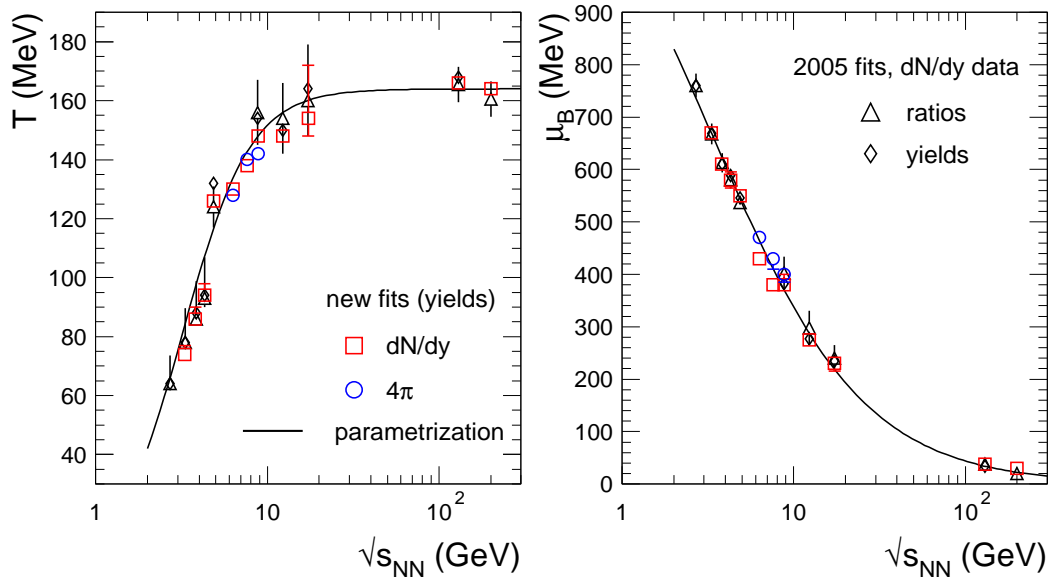}}
\end{center}
\caption{The temperature and baryon chemical potential of Statistical Model
 fits to hadro-chemical abundances as a function of center of mass energy per
 nucleon pair for collisions of heavy nuclei (Figure taken from
 ~\cite{Andronic:2008gu,Andronic:2009qf}).}
\label{energy_t_mu}
\end{figure}

These data have been analyzed independently by various groups, using hadron
resonance gas models with chemical freeze-out as a key ingredient. In this
resonance gas approach, described in detail in \cite{hwa}, it is assumed that
the final hadron yields result from the decay of fully equilibrated hadronic
matter, comprising the full QCD hadron mass spectrum - and thereby implicitely
containing all of QCD. The measured hadrons, those not decaying further by the
strong interaction, are then obtained by summing up all strong decay products
of the hadron mass spectrum with their corresponding thermal weight.  By
assuming that the measured hadrons are generated at a common surface at which
all particles simultaneously decouple, values of the baryon chemical
potential, $\mu_B$, and temperature, $T$, on this surface, the chemical
freeze-out surface, are extracted.  Fitting these two parameters, $\mu_b$ and
$T$, together with the volume parameter gives values for the particle
abundances which are in very good agreement with experiment
\cite{BraunMunzinger:1994xr,BraunMunzinger:1995xr,Cleymans:1999st,Heinz:1999kb,
  Letessier:2000ay,BraunMunzinger:2001ip,Florkowski:2001fp,Becattini:2005xt,Becattini_n,
  Cleymans:2005zx,Andronic:2005yp,corw,Andronic:2008gu,Andronic:2009qf}, for
an overall summary see \cite{hwa}.

These results clearly demonstrate that, at chemical freeze-out, a very high
degree of equilibration is reached in central nucleus-nucleus collisions,
implying that matter in the proper sense with thermodynamic properties has been formed at the
latest at this stage of the collision. We note that the system most likely is
in or very close to equilibrium much earlier in the collision, as demonstrated
by, among other things, hydrodynamic analysis of azimuthal anisotropies
('elliptic flow'). For a brief summary see \cite{BraunMunzinger:2007zz,gyulassy_mclerran}.

The resulting values of $\mu_B$ and $T$ are shown in Fig.~\ref{energy_t_mu} as
functions of center-of-mass energy per nucleon pair.
We note first a very important outcome of these investigations: near 10 GeV
center of mass energy, the temperature saturates with increasing beam energy,
reaching an asymptotic value of about 160 MeV, while the baryon chemical
potential decreases smoothly. This temperature saturation concept can be
tested thoroughly when the newly taken data at the much higher energy of
$\sqrt{s_{NN}} = 2.76$ TeV from the LHC are analyzed.

All these data points have been obtained by analyzing hadron yields,
implying the existence of a limiting temperature to which a hadron
'gas' can be heated, as has been conjectured more than 40 years ago by
Hagedorn \cite{Hage}. The constancy of $T$ above $\sqrt{s_{NN}} = 10$
GeV indicates that a boundary has been reached. Is this the QCD phase
boundary? Could there be hadronic matter hotter than is observed here?
Since the data analyzed here comprise the full hadronic spectrum, this
would imply the existence of a hot and dense hadronic medium between
the QCD phase boundary and the chemical freeze-out curve that leaves
no trace in hadronic observables. As we will discuss below, this would
contradict the existence of chemical freeze-out outlined above.

\begin{figure}[htb]
\begin{center}
\resizebox*{!}{10.0cm}{\includegraphics{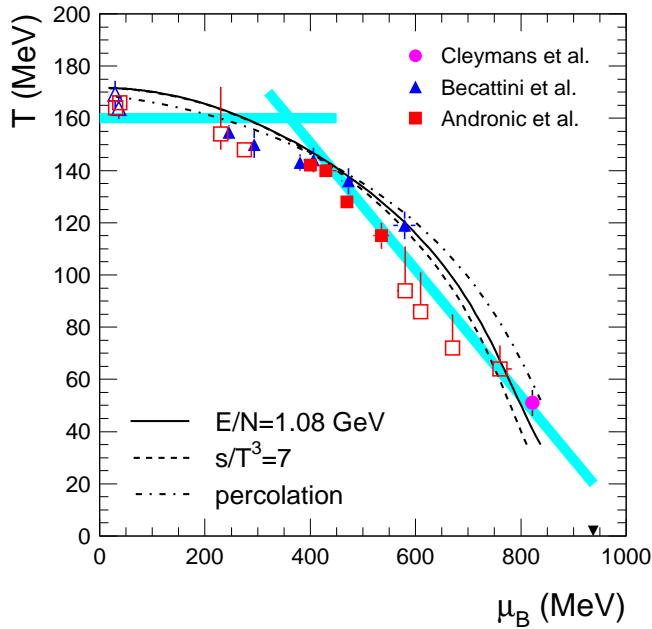}}
\end{center}
\caption{The decoupling temperatures and chemical potentials extracted by
  Statistical Model fits to experimental data. The freeze-out points are from
  Refs. \cite{Becattini:2005xt} and \cite{Andronic:2008gu,fr1,fr2}. The open
  points are obtained from fits to data at mid-rapidity, the full-points refer
  to $4\pi$ data. The inverse triangle at $T=0$ indicates the position of
  normal nuclear matter at $T=0$. The lines are different model calculations
  to provide a phenomenological understanding of the freeze-out curve
  \cite{corw,per,1gev}. The shaded lines are drawn to indicate different
  regimes in this diagram. Figure taken from \cite{quarkyonic}}
\label{line}
\end{figure}

Plotting these temperature-chemical potential pairs for all available
energies results in a phase diagram-like picture as is illustrated in
Fig. \ref{line}.  In the $\mu_B$ region from 800 to 400 MeV, as $T$
increases from 50 to 150 MeV, the experimental points rise
approximately linearly. In contrast, below $\mu_B \simeq 400$~MeV, the
temperature is approximately constant, $T \simeq 160$~MeV. The highest
collision energies studied to date at RHIC are those for which
$\mu_B\simeq 25$~MeV. Soon an experimental point at $\mu_B \simeq
1$~MeV will be available from the recent campaign at the LHC.  Also
shown on this plot are lines of fixed energy per particle and fixed
entropy density per $T^3$ as well as a line of hadron percolation
\cite{corw,per,1gev}.

These experimental results can be compared to the phase boundary
computed on the lattice \cite{lattice_review,ghk}.  Numerical
simulations in lattice QCD can be performed at nonzero temperature
only for small values of $\mu_B$, typically $\mu_B < T$, without
running into the 'sign' problem. At $\mu_B = 0$, these simulations
indicate that there is no true phase transition from hadronic matter
to a quark-gluon plasma, but rather a very rapid rise in the energy
density at a temperature $T_c$ of $155-175$~MeV within current
systematic errors, in very close agreement with the chemical
freeze-out temperatures determined for small values of $\mu_B$.

Further, studies using the lattice technique imply that $T_c$
decreases very little as $\mu_B$ increases, at least for moderate
values of $\mu_B$. A very recent study \cite{karsch2010} provides a
quantitative estimate of the curvature of the phase boundary near
$\mu_B=0$. By comparison to the curvature of the line of fixed
energy/particle \cite{1gev} which has been constructed to get a
phenomenological understanding of the complete chemical freeze-out
curve, these authors argue that there is a significant difference
between the data on chemical freeze-out and the lattice predictions,
implying in their view the existence of a dense hadronic phase between
chemical freeze-out and the QCD phase boundary.

\begin{figure}[htb]
\begin{center}
\resizebox*{!}{9.0cm}{\includegraphics{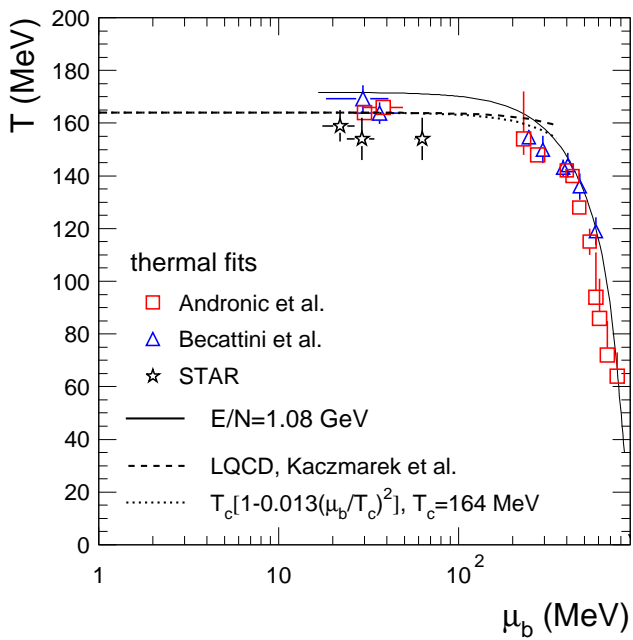}}
\end{center}
\caption{Comparison of the experimental $\mu_B$ dependence of the chemical
  freeze-out temperature with recent lattice predictions
  \cite{karsch2010}. The STAR chemical freeze-out values are from
  \cite{star_chem}, the other values 
  are from \cite{quarkyonic}. For more discussion see text. 
}
\label{slope}
\end{figure}

A precise determination of the curvature of the chemical freeze-out curve can,
however, only be obtained by analyzing the data themselves, not a
curve such as the line of fixed energy/particle or any of the
other lines shown in Fig. \ref{line}. We have recently investigated
\cite{andronic2010} the curvature of the chemical freeze-out curve in the
region $\mu_B < 250$ MeV. The results are presented in Fig.~\ref{slope}. The
chemical freeze-out points are taken from \cite{quarkyonic}. In addition, we
show in Fig.~\ref{slope} also fit results from the STAR collaboration
\cite{star_chem}, where only pions, kaons and protons are analyzed to
determine the freeze-out curve. This leads to a slightly lower chemical
freeze-out temperature, but we believe that its energy (or $\mu_B$) dependence
contains the correct information. Independent of which chemical freeze-out
parameters are used, the
resulting  description using the  parabolic fit function as in
\cite{karsch2010} (long dashes) and the same  function with twice the slope parameter
(short dashes)
are in agreement with the data, within the (still significant
systematic) uncertainties, as shown in Fig.~\ref{slope}. For reference we
include, in Fig.~\ref{slope}, the line of constant E/N \cite{1gev}. From the available
data one cannot construct evidence for a dense hadronic phase between chemical
freeze-out and the phase boundary.

\begin{figure}[htb]
\begin{tabular}{lr} \begin{minipage}{.49\textwidth}
\hspace{-.3cm}\includegraphics[width=1.\textwidth]{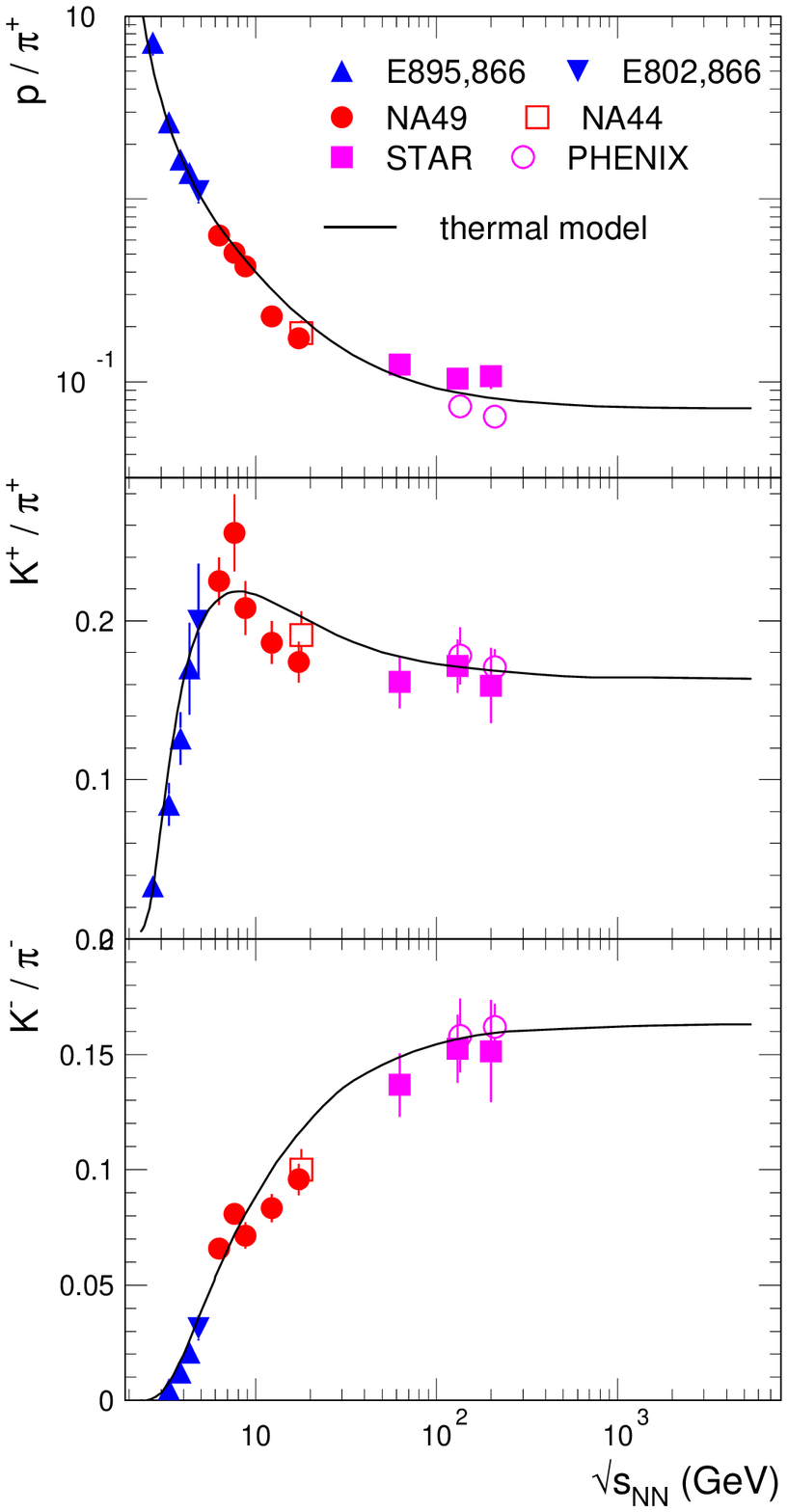}
\end{minipage} &\begin{minipage}{.49\textwidth}
\hspace{-.5cm}\includegraphics[width=1.\textwidth]{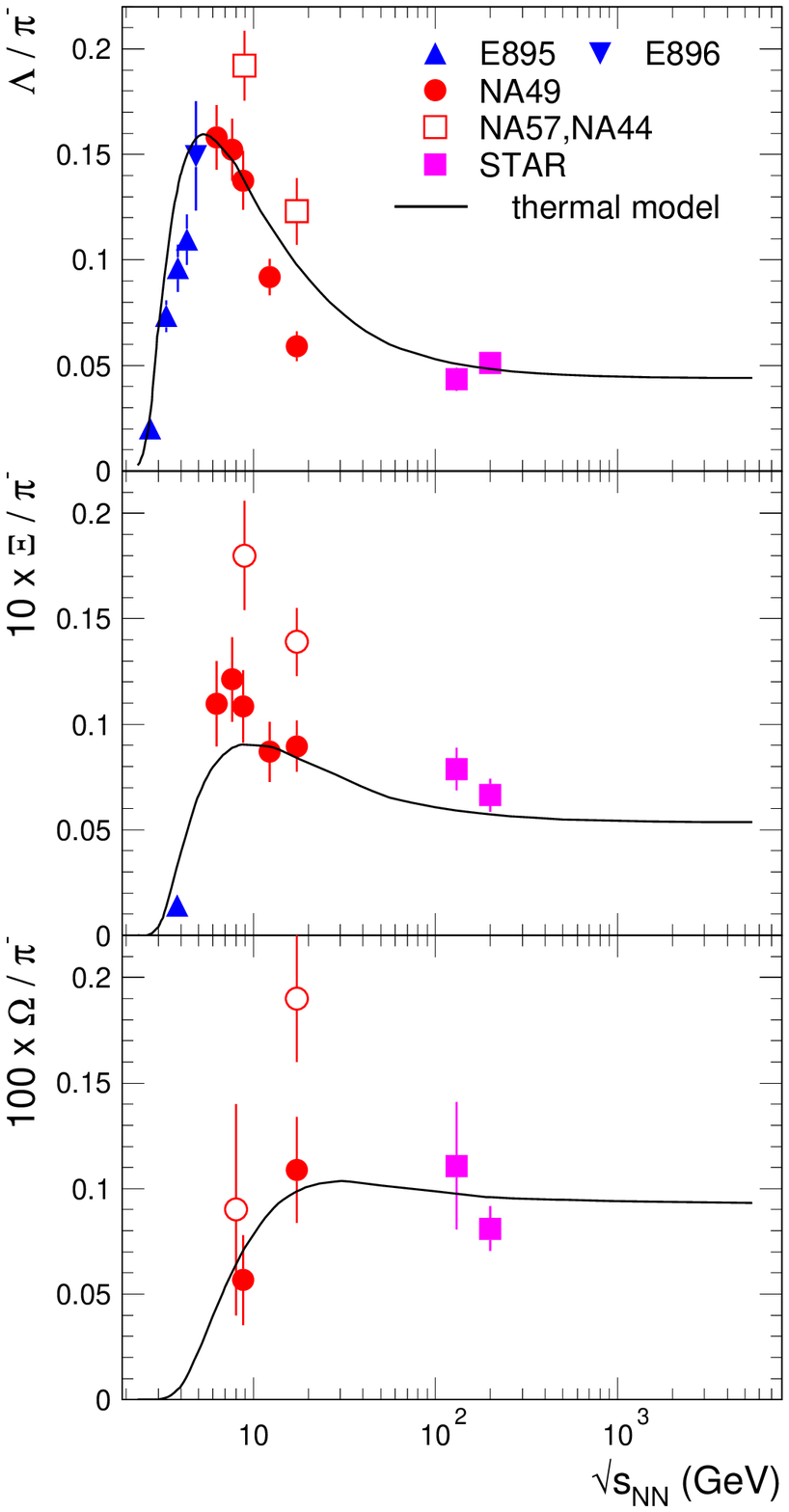}
\end{minipage} \end{tabular}
\caption{Energy dependence of hadron yields relative to pions. The points are
  experimental data from various experiments.  Lines are results of the
  Statistical Model calculations. The Figure is taken from
  ~\cite{Andronic:2008gu,Andronic:2009qf}).}
\label{fig_k2pi}
\end{figure}

\begin{figure}[htb]
\begin{center}
\resizebox*{!}{9.0cm}{\includegraphics{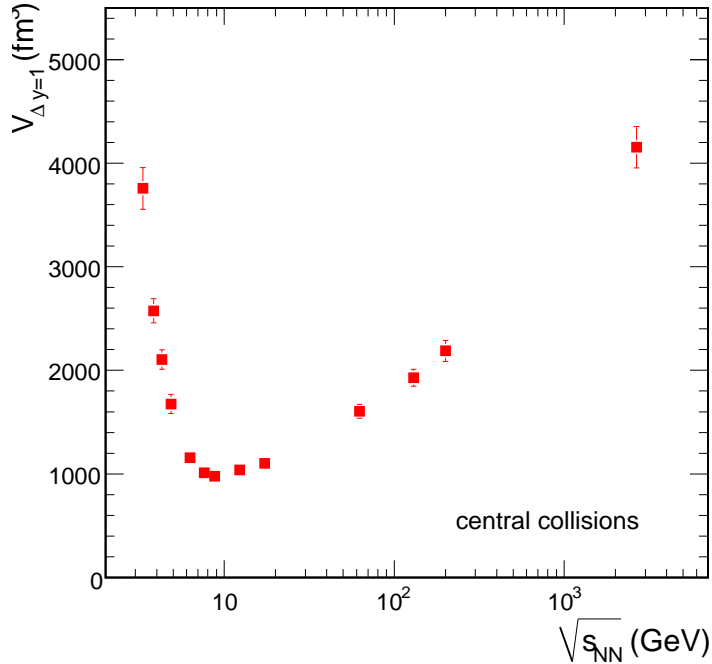}}
\end{center}
\caption{Energy dependence of the volume for central nucleus-nucleus
  collisions. The chemical freeze-out volume dV/dy for one unit of rapidity
  is taken from Ref. \cite{Andronic:2005yp}.  The LHC point is from
  analysis of the recent ALICE data \cite{alice_pb1}.
}
\label{size}
\end{figure}

With the parametrizations of $T$ and $\mu_B$ from
Fig.~\ref{energy_t_mu} one can compute the energy dependence of the
production yields of various hadrons relative to pions, shown in
Fig.~\ref{fig_k2pi}.  Important for our purposes is the observation
that there are peaks in the abundances of strange to non-strange
particles at center of mass energies near 10 GeV, i.e. where the
temperature reaches its limiting value. In particular, the $K^+/\pi^+$
and $\Lambda/\pi$ ratios exhibit rather pronounced maxima there.  We
further note that in the region near 10 GeV, there is also a minimum
in the chemical freeze-out volume \cite{Andronic:2005yp,toneev}
obtained from the Statistical Model fit to particle yields
\cite{Andronic:2005yp,Andronic:2009qf}, as well as in the thermal freeze-out volume
obtained from the Hanbury-Brown and Twiss (HBT) radii of the fireball
\cite{Adamova:2002ff}.  The energy dependence of the chemical freeze-out volume
is shown in Fig. \ref{size}. Included in this figure is the most recent point
at LHC energy from ALICE \cite{alice_pb1} obtained by analyzing the
pseudo-rapidity density of charged particles for central Pb--Pb collisions.

These experimental observations have long resisted interpretation in terms of
a transition between hadronic matter and a quark-gluon plasma\footnote{We note
  the interpretation given in \cite{Gazdzicki:1998vd}, obtained within a
  schematic 1st order phase transition model.}. The general structures
observed in the data are well reproduced only by the most recent model
calculations \cite{Andronic:2008gu}. There, it is shown that these structures
arise due to the interplay between the limit in hadronic temperature (see
Fig.~\ref{energy_t_mu}), in our interpretation due to the QCD phase
transition, and the rapid decrease of $\mu_B$ with increasing energy. The
minimum near a center of mass energy of 10 GeV and the rise towards higher
energies in the thermal freeze-out volume \cite{Adamova:2002ff} was explained
as due to the increasing meson to baryon ratio in this region combined with a
meson-meson cross section which is significantly smaller than the pion-nucleon
cross section. The chemical freeze-out volume depicted in Fig.~\ref{size}
behaves similarily, with a minimum at an energy close to where the chemical
freeze-out temperature saturates. At the highest (LHC) energy the chemical
freeze-out volume reaches, for Pb--Pb central collisions, a value of about 4
times, the thermal freeze-out volume about 7 times that of a Pb nucleus. The
strong volume increase between RHIC and LHC energy, where neither $T$ nor
$\mu_B$ are expected to change much, provides a strong argument that chemical
freeze-out takes place at a fixed density.  These findings further strengthen
the connection between hadron gas and quark-gluon plasma.

The same chemical freeze-out mechanism also governs the production of
light nuclei and their anti-particles and even complex and exotic
objects such as light hypernuclei and their anti-particles
\cite{pbm_com95,pbm_com2001,andronic_com2009}.  This comes at first
glance as a surprize to many, as the binding energy of these objects
is typically 1-2 orders of magnitude smaller than the chemical
freeze-out temperature of $T \approx 160$~MeV. Can it be that such
loosely bound objects are messengers of the phase transition between
hadronic matter and the quark-gluon plasma?  Entropy conservation
comes to the rescue! After the phase transition the hot fireball
expands adiabatically and the overall entropy as well as the
entropy/baryon are conserved quantities. As was realized already more
than 30 years ago \cite{siemens_kapusta} the assembly yield of complex
nuclei from a gas of hot nucleons is a measure of the entropy/baryon,
and hence can be used to diagnose the hot and dense phase of the
collision. The answer to the above question is therefore strongly
affirmative: the yield of composite objects such as light nuclei and
hyper-nuclei (and their anti-particles) produced in central
nucleus-nucleus collisions at ultra-rlativistic energy is a direct
measure of the entropy/baryon at chemical freeze-out, and hence, of
the QCD phase transition.

At RHIC energies, chemical freeze-out was shown
\cite{BraunMunzinger:2003zz} to take place very close (within less
than about 10 MeV) to the phase boundary, driven by multi-particle
collisions in the high density regime of, and by the the rapid density
change across the phase transition. Further it is argued that
freeze-out ends when the system is fully hadronized, i.e. at
comparatively low density in the hadronic phase.  Without a phase
transition, freeze-out in a purely hadronic medium takes place over a
considerable time and temperature range, as has been recently
demonstrated from very general considerations \cite{knol}. This would
necessarily lead to different freeze-out parameters for each hadron
species due to widely different hadronic cross sections. This is not
observed. Conversely, the rapid change of density with temperature
near the phase transition makes the system insensitive to the
different cross sections. We conclude that the observed (nearly)
simultaneous chemical freeze-out of all hadrons is inconsistent with
hypothesis of a dense hadronic phase between the chemical freeze-out
line and the phase boundary, at least for relatively small values of
$\mu_B < 250$~MeV.

We believe, however, that the above argument is generic
\cite{BraunMunzinger:2003zz}: to ensure simultaneous freeze-out (within a very
small interval in temperature and chemical potential)  of
all hadrons, the freeze-out curve has to be very close to a line with
a rapid density change, significantly faster than can be obtained
within the scenario of an expanding hadronic medium.  An immediate
consequence of this would be that the chemical freeze-out curve
delineates phase boundaries, not only for small values of $\mu_B$ but
everywhere.

But what provides the phase boundary for large values of $\mu_B$,
where the deconfinement transition seems far away, at least if one
follows the guidance from lattice QCD calculations?  In line with
Gerry Brown's daring attitude concerning new scientific directions we
have recently speculated \cite{quarkyonic} that the transition from
hadronic to quarkyonic matter might provide the missing link. Another,
related possibility might be that, at high baryon density, the lines
corresponding to the chiral and deconfinement phase transition split,
leading to a region of chirally symmetric but still confined matter in
the QCD phase diagram.

Whatever the case may be, it is clear that the phase diagram at high
baryon density is difficult to explore, both experimentally and
theoretically. Speculations have been put forward about the possible
existence of a critical endpoint \cite{stephanov} or a triple
point \cite{quarkyonic} in the QCD phase diagram. However, despite
significant efforts, no clear signals have been discovered to-date
\cite{cbm_physicsbook,rhic_lowenergy} of either critical points or,
more generally, of dense phases beyond chemical freeze-out. This
research on the high baryon density region of the phase diagram
remains a challenge for the future and will have to be largely driven
by the running campaign at RHIC \cite{rhic_lowenergy} and the planned
new experiments at FAIR \cite{stoecker} and DUBNA \cite{sissakian} .
  
%\section{Conclusions}

We have provided strong evidence that, for not too large values of the
baryo-chemical potential $\mu_B < 250$~MeV, the experimentally observed
chemical freeze-out curve closely coincides with the QCD phase boundary
between the hadronic world and the quark-gluon plasma. This implies a direct
connection between results from experiments with ultra-relativistic nuclear
collisions and a fundamental prediction of QCD concerning the phase structure
of strongly interacting matter. For large values of $\mu_B$, i.e. large net
baryon densities, we can presently only speculate, but in Gerry Brown fashion
have put forward an argument that also there the chemical freeze-out curve is
driven by a phase transition. 

%%%%%%%%%%%%%%%%%%%%%%%%%%%%%%%%%%%%%
\section*{Acknowledgments}
The authors would like to thank Anton Andronic for many important
contributions on the subject discussed.  
%%%%%%%%%%%%%%%%%%%%%%%%%%%%%%%


\begin{thebibliography}{99}

\bibitem{pbm_wambach}
  P.~Braun-Munzinger and J.~Wambach,
  %``Colloquium: Phase diagram of strongly interacting matter,''
  Rev.\ Mod.\ Phys.\  {\bf 81}, 1031 (2009).
 
\bibitem{fukushima_hatsuda}
  K.~Fukushima and T.~Hatsuda,
  %``The phase diagram of dense QCD,''
  Rept.\ Prog.\ Phys.\  {\bf 74}, 014001 (2011);
  arXiv:1005.4814 [hep-ph].
  
\bibitem{pbm_js_gerry70}
  P.~Braun-Munzinger and J.~Stachel,
  %``Probing the phase boundary between hadronic matter and the
  %quark-gluon-plasma in relativistic heavy ion collisions,''
  Nucl.\ Phys.\  A {\bf 606}, 320 (1996);
  arXiv:nucl-th/9606017.
 
\bibitem{welke} G.E. Brown, J. Stachel, and G. Welke, Phys. Lett. B {\bf
  253}, 19 (1991).

\bibitem{Specht:2001qe}
 H.~J.~Specht,
 %``Experimental conference summary,''
 Nucl.\ Phys.\  A {\bf 698}, 341c (2002).
% [arXiv:nucl-ex/0111011].
 %%CITATION = NUPHA,A698,341;%%

\bibitem{Heinz:2000bk}
U.~W.~Heinz and M.~Jacob,
%``Evidence for a new state of matter: An assessment of the results from
%the %CERN lead beam programme,''
arXiv:nucl-th/0002042.
%%CITATION =NUCL-TH/0002042;%%

\bibitem{Stachel:1998rc}
 J.~Stachel,
 %``Towards the quark-gluon plasma,''
 Nucl.\ Phys.\  A {\bf 654}, 119c (1999).
% [arXiv:nucl-ex/9903007].
 %%CITATION = NUPHA,A654,119C;%%

\bibitem{Arsene:2004fa}
I.~Arsene {\it et al.} [BRAHMS Collaboration],
%``Quark Gluon Plasma an Color Glass Condensate at
%RHIC? The perspective from %the BRAHMS experiment,''
Nucl.\ Phys.\  A {\bf 757}, 1 (2005).
%[arXiv:nucl-ex/0410020].
%%CITATION = NUPHA,A757,1;%%

\bibitem{Back:2004je}
B.~B.~Back {\it et al.} [PHOBOS Collaboration],
%``The PHOBOS perspective on discoveries at RHIC,''
Nucl.\ Phys.\  A {\bf 757}, 28 (2005).
%[arXiv:nucl-ex/0410022].
%%CITATION = NUPHA,A757,28;%%

\bibitem{Adams:2005dq}
J.~Adams {\it et al.}  [STAR Collaboration],
%``Experimental and theoretical challenges in the search for the quark  gluon
%plasma: The STAR collaboration's critical assessment
%of the  evidence from %RHIC collisions,''
Nucl.\ Phys.\  A {\bf 757}, 102 (2005).
%[arXiv:nucl-ex/0501009].
%%CITATION = NUPHA,A757,102;%%

\bibitem{Adcox:2004mh}
K.~Adcox {\it et al.}  [PHENIX Collaboration],
%``Formation of dense partonic matter in relativistic
%nucleus nucleus %collisions at RHIC: Experimental evaluation by the PHENIX
%collaboration,''
Nucl.\ Phys.\  A {\bf 757}, 184 (2005).
%[arXiv:nucl-ex/0410003].
%%CITATION = NUPHA,A757,184;%%

\bibitem{BraunMunzinger:2007zz}
 P.~Braun-Munzinger and J.~Stachel,
 %``The Quest For The Quark-Gluon Plasma,''
 Nature {\bf 448}, 302 (2007).
 %%CITATION = NATUA,448,302;%%

\bibitem{alice_pb1}
  K.~Aamodt {\it et al.}  [ALICE Collaboration], Phys. Rev. Lett., in
  print,
  %``Charged-particle multiplicity density at mid-rapidity in central Pb-Pb
  %collisions at sqrt(sNN) = 2.76 TeV,''
  arXiv:1011.3916 [nucl-ex].
 
\bibitem{alice_pb2}
  K.~Aamodt {\it et al.}  [ALICE Collaboration], Phys. Rev. Lett., in print,
  %``Elliptic flow of charged particles in Pb-Pb collisions at 2.76 TeV,''
  arXiv:1011.3914 [nucl-ex].

\bibitem{alice_pb3}
  K.~Aamodt {\it et al.}  [ALICE Collaboration], Phys. Lett. B, in print,
  %``Suppression of Charged Particle Production at Large Transverse
  %Momentum in central Pb--Pb Collisions at $\sqrt{s_{_{NN}}} = 2.76$
  %TeV,''
  arXiv:1012.1004 [nucl-ex].

\bibitem{alice_pb4}
  K.~Aamodt {\it et al.}  [ALICE Collaboration],
  %``Centrality dependence of the charged particle multiplicity at
  %mid-rapidity in
  %Pb--Pb Collisions at $\sqrt{s_{_{NN}}} = 2.76$ TeV,''
  arXiv:1012.1657 [nucl-ex].
 
\bibitem{alice_pb5}
   K.~Aamodt {\it et al.}  [ALICE Collaboration],
  %``Two-pion Bose-Einstein correlations in central PbPb collisions at
  %sqrt(s_NN) = 2.76 TeV,''
  arXiv:1012.4035 [nucl-ex].
 
\bibitem{atlas} G. Aad {\it et al.} [ATLAS Collaboration],
Phys. Rev. Lett. {\bf 105} (2010) 252203. 

\bibitem{gyulassy_mclerran}
  M.~Gyulassy and L.~McLerran,
  %``New forms of QCD matter discovered at RHIC,''
  Nucl.\ Phys.\  A {\bf 750}, 30 (2005)
  [arXiv:nucl-th/0405013].


\bibitem{hwa}
P. Braun-Munzinger, K. Redlich, and J. Stachel, in
Quark-Gluon Plasma 3, Eds. R.C. Hwa and X.N. Wang,
(World Scientific Publishing, 2004) 491; nucl-th/0304013.

\bibitem{BraunMunzinger:1994xr}
 P.~Braun-Munzinger, J.~Stachel, J.~P.~Wessels, and N.~Xu,
 %``Thermal equilibration and expansion in nucleus-nucleus collisions at the
 %AGS,''
 Phys.\ Lett.\  B {\bf 344}, 43 (1995).
% [arXiv:nucl-th/9410026].
%%CITATION = PHLTA,B344,43;%%

\bibitem{BraunMunzinger:1995xr}
 P.~Braun-Munzinger, J.~Stachel, J.~P.~Wessels, and N.~Xu,
 Phys.\ Lett.\  B {\bf 365}, 1 (1996).

\bibitem{Cleymans:1999st}
 J.~Cleymans and K.~Redlich,
 %``Chemical and thermal freeze-out parameters from 1-A-GeV to 200-A-GeV,''
 Phys.\ Rev.\  C {\bf 60}, 054908 (1999).
% [arXiv:nucl-th/9903063].
%%CITATION = PHRVA,C60,054908;%%

\bibitem{Heinz:1999kb}
 U.~W.~Heinz,
 %``Primordial hadrosynthesis in the little bang,''
 Nucl.\ Phys.\  A {\bf 661}, 140 (1999).
% [arXiv:nucl-th/9907060].
 %%CITATION = NUPHA,A661,140;%%

\bibitem{Letessier:2000ay}
 J.~Letessier and J.~Rafelski,
 %``Observing quark-gluon plasma with strange hadrons,''
 Int.\ J.\ Mod.\ Phys.\  E {\bf 9}, 107 (2000).
% [arXiv:nucl-th/0003014].
 %%CITATION = IMPAE,E9,107;%%

\bibitem{BraunMunzinger:2001ip}
 P.~Braun-Munzinger, D.~Magestro, K.~Redlich, and J.~Stachel,
 %``Hadron production in Au Au collisions at RHIC,''
 Phys.\ Lett.\  B {\bf 518}, 41 (2001).
% [arXiv:hep-ph/0105229].
 %%CITATION = PHLTA,B518,41;%%

\bibitem{Florkowski:2001fp}
 W.~Florkowski, W.~Broniowski, and M.~Michalec,
%``Thermal analysis of particle ratios and p(T) spectra at RHIC,''
Acta Phys.\ Polon.\  B {\bf 33}, 761 (2002).
%[arXiv:nucl-th/0106009].
 %%CITATION = APPOA,B33,761;%%

\bibitem{Becattini:2005xt}
 F.~Becattini, J.~Manninen, and M.~Ga\'zdzicki,
 %``Energy and system size dependence of chemical freeze-out in  relativistic
 %nuclear collisions,''
 Phys.\ Rev.\  C {\bf 73}, 044905 (2006).
%[arXiv:hep-ph/0511092].
 %%CITATION = PHRVA,C73,044905;%%

\bibitem{Becattini_n}
F. Becattini {\it et al.},
Phys. Rev. C {\bf 64}, 024901 (2001).
%[hep-ph/0002267]

\bibitem{Cleymans:2005zx}
 J.~Cleymans, H.~Oeschler, K.~Redlich,  and S.~Wheaton,
 %``The thermal model and the transition from baryonic to mesonic
 %freeze-out,''
 Eur.\ Phys.\ J.\  A {\bf 29}, 119 (2006)
%[arXiv:hep-ph/0510283].
 %%CITATION = EPHJA,A29,119;%%

\bibitem{Andronic:2005yp}
A.~Andronic, P.~Braun-Munzinger, and J.~Stachel,
%``Hadron production in central nucleus nucleus collisions at chemical
%freeze-out,''
Nucl.\ Phys.\  A {\bf 772}, 167 (2006).
%[arXiv:nucl-th/0511071].
%%CITATION = NUPHA,A772,167;%%

\bibitem{corw}
%`Comparison of Chemical Freeze-Out Criteria in Heavy-Ion Collisions.''\\
J. Cleymans, H. Oeschler, K. Redlich, and S. Wheaton,
Phys. Rev. C {\bf 73}, 034905 (2006).

\bibitem{Andronic:2008gu}
A.~Andronic, P.~Braun-Munzinger, and J.~Stachel,
%``Thermal hadron production in relativistic nuclear collisions: the sigma
%meson, the horn, and the QCD phase transition,''
Phys.\ Lett.\  B {\bf 673}, 142 (2009);
%[arXiv:0812.1186[nucl-th]].
%%CITATION = PHLTA,B673,142;%%
 Erratum, {\it ibid.} B {\bf 678}, 516 (2009).

\bibitem{Andronic:2009qf}
A.~Andronic, P.~Braun-Munzinger, and J.~Stachel,
%``Thermal hadron production in relativistic nuclear collisions,''
Acta Phys.\ Polon.\  B {\bf 40}, 1005 (2009).
%[arXiv:0901.2909 [nucl-th]].
 %%CITATION = APPOA,B40,1005;%%

\bibitem{Hage} R.\ Hagedorn, Nuovo Cim. Suppl. {\bf 3},  147 (1965);
Nuovo Cim. A {\bf 56}, 1027 (1968).

\bibitem{fr1}
J. Manninen and F. Becattini, Phys. Rev. C {\bf 78}, 054901 (2008).
%[arXiv:0806.4100[nucl-th]],

\bibitem{fr2}
J. Cleymans et al., Phys. Rev. C {\bf 59}, 1663 (1999).
%[nucl-th/9809027].


\bibitem{per}
V. Magas and H. Satz, Eur. Phys. J. C {\bf 32}, 115 (2003).

\bibitem{1gev}
J. Cleymans  and  K. Redlich,  Phys. Rev. Lett. {\bf 81}, 5284 (1998).
%[nucl-th/9808030].

\bibitem{quarkyonic}
  A.~Andronic {\it et al.},
  %``Hadron Production in Ultra-relativistic Nuclear Collisions: Quarkyonic
  %Matter and a Triple Point in the Phase Diagram of QCD,''
  Nucl.\ Phys.\  A {\bf 837}, 65 (2010);
  arXiv:0911.4806 [hep-ph].


\bibitem{lattice_review}
C.~DeTar and U.~M.~Heller,
%``QCD Thermodynamics from the Lattice,''
Eur. Phys. J. A {\bf 41}, 405 (2009).
%[arXiv:0905.2949 [hep-lat]].
%%CITATION = ARXIV:0905.2949;%%

\bibitem{ghk}
Y.~Aoki, Z.~Fodor, S.~D.~Katz, and K.~K.~Szabo,
%``The QCD transition temperature: Results with physical masses in the
%continuum limit,''
\plb{643}{46}{2006};
%\href{http://xxx.lanl.gov/abs/hep-lat/0609068}{[arXiv:hep-lat/0609068]}.
%%CITATION = PHLTA,B643,46;%%
M.~Cheng {\it et al.},
%``The QCD Equation of State with almost Physical Quark Masses,''
\prd{77}{014511}{2008};
%\href{http://arXiv.org/abs/0710.0354}{[arXiv:0710.0354]};
%%CITATION = PHRVA,D77,014511;%%
A.~Bazavov {\it et al.},
%``Equation of state and QCD transition at finite temperature,''
%\href{http://xxx.lanl.gov/abs/0903.4379}{[arXiv:0903.4379]}.
\prd{80}{014504}{2009}.
%[arXiv:0903.4379].
%%CITATION = ARXIV:0903.4379;%%
%%CITATION = APPOA,B30,2705;%%

\bibitem{karsch2010}
  F.~Karsch {\it et al.},
  %``The phase boundary for the chiral transition in (2+1)-flavor QCD at small
  %values of the chemical potential,''
  arXiv:1011.3130 [hep-lat].

\bibitem{andronic2010} A. Andronic, P. Braun-Munzinger and J. Stachel,
manuscript in preparation.

\bibitem{star_chem} B.~I. Abelev, {\it et al.}, [STAR Collaboration],
Phys. Rev. {\bf C79} 034909 (2009). 

\bibitem{toneev} V.~D.~Toneev and A.~S.~Parvan, J. Phys. G {\bf 31}, 583 (2005).
% Canonical Strangeness and Distillation Effects in Hadron Production
%[arXiv:nucl-th/0411125]

\bibitem{Adamova:2002ff}
D.~Adamova {\it et al.}  [CERES Collaboration],
%``Universal pion freeze-out in heavy-ion collisions,''
Phys.\ Rev.\ Lett.\  {\bf 90}, 022301 (2003).
%[arXiv:nucl-ex/0207008].
%%CITATION = PRLTA,90,022301;%%

\bibitem{Gazdzicki:1998vd}
M.~Gazdzicki and M.~I.~Gorenstein,
%``On the early stage of nucleus nucleus collisions,''
Acta Phys.\ Polon.\  B {\bf 30}, 2705 (1999).
%[arXiv:hep-ph/9803462].

\bibitem{pbm_com95} P. Braun-Munzinger and J. Stachel,
% Production of Strange Clusters
%and Strange Matter in Nucleus-Nucleus Collisions at the AGS,
J. Phys. G {\bf 21}, L17 (1995).

\bibitem{pbm_com2001} P. Braun-Munzinger and J. Stachel, 
%Particle Ratios, Equilibration,
%and the QCD Phase Boundary, Contributed to 6th International
%Conference on Strange Quarks in Matter: 2001: A Flavorspace Odyssey
%(SQM2001), Frankfurt, Germany, 25-29 Sep 2001. 
J. Phys. G {\bf 28}, 1971 (2002), nucl-th/0112051. 

\bibitem{andronic_com2009}  A. Andronic, P. Braun-Munzinger,
  J. Stachel and  H. St\"ocker,
% Production of
%  light nuclei, hypernuclei and their antiparticles in relativistic nuclear
%  collisions, 
Phys. Lett. B submitted Oct. 15, 2010, arXiv:1010.2995 [nucl-th].

\bibitem{siemens_kapusta}
  P.~J.~Siemens and J.~I.~Kapusta,
  %``Evidence For A Soft Nuclear Matter Equation Of State,''
  Phys.\ Rev.\ Lett.\  {\bf 43}, 1486 (1979).
 
\bibitem{BraunMunzinger:2003zz}
P.~Braun-Munzinger, J.~Stachel, and C.~Wetterich,
%``Chemical freeze-out and the QCD phase transition temperature,''
Phys.\ Lett.\  B {\bf 596}, 61 (2004).
%[arXiv:nucl-th/0311005].
%%CITATION = PHLTA,B596,61;%%

\bibitem{knol}
J. Knoll, Nucl. Phys. A {\bf 821} 235 (2009).
  
\bibitem{stephanov}
M.~A.~Stephanov, K.~Rajagopal, and E.~V.~Shuryak,
%``Signatures of the tricritical point in {QCD},''
Phys.\ Rev.\ Lett.\  {\bf 81}, 4816 (1998);
%[arXiv:hep-ph/9806219];
%%CITATION = PRLTA,81,4816;%%
%``Event-by-event fluctuations in heavy ion collisions and the {QCD}  critical
%point,''
Phys.\ Rev.\  D {\bf 60}, 114028 (1999).
%[arXiv:hep-ph/9903292].
%%CITATION = PHRVA,D60,114028;%%

\bibitem{cbm_physicsbook} The CBM Physics Book: Compressed Baryonic Matter in
  Laboratory Experiments, B. Friman, C. Hoehne, J. Knoll, S. Leupold,
  J. Randrup, R. Rapp, P. Senger, editors, Lecture Notes In Physics,
  (Springer, Heidelberg, 2010).

\bibitem{rhic_lowenergy}
  M.~M.~Aggarwal {\it et al.}  [STAR Collaboration],
  %``Higher Moments of Net-proton Multiplicity Distributions at RHIC,''
  Phys.\ Rev.\ Lett.\  {\bf 105}, 022302 (2010);
  arXiv:1004.4959 [nucl-ex].

\bibitem{stoecker} J.~Steinheimer, H.~St\"ocker, I.~Augustin,
  A.~Andronic, T.~Saito and P.~Senger,
  %``Strangeness at the international Facility for Antiproton and Ion
  %Research,''
  Prog.\ Part.\ Nucl.\ Phys.\  {\bf 62}, 313 (2009).

\bibitem{sissakian}
  A.~N.~Sissakian and A.~S.~Sorin  [NICA Collaboration],
  %``The nuclotron-based ion collider facility (NICA) at JINR: New prospects for
  %heavy ion collisions and spin physics,''
  J.\ Phys.\ G {\bf 36}, 064069 (2009).

\end{thebibliography}
\end{document}